\documentclass[aps,prl,twocolumn,superscriptaddress,nofootinbib]{revtex4-2}

\usepackage{amsmath,amssymb,bm,graphicx,color}
\usepackage[colorlinks=true,citecolor=blue,linkcolor=blue,urlcolor=blue]{hyperref}
\usepackage{braket}
\usepackage{booktabs,array,needspace,microtype}

\newcommand{\blue}[1]{{\color{blue}{#1}}}

\newcommand{\ii}{i}

\begin{document}

\title{Unified First-Principles Theory of Finite-Wave-Vector Photocurrents in Crystals}

\author{Bumseop Kim}
\affiliation{Department of Chemistry, University of Pennsylvania, Philadelphia, Pennsylvania 19104, USA}

\author{Sayed Ali Akbar Ghorashi}
\affiliation{Department of Chemistry, University of Pennsylvania, Philadelphia, Pennsylvania 19104, USA}

\author{Yicong Chen}
\affiliation{Department of Physics and Astronomy, University of Pennsylvania, Philadelphia, Pennsylvania 19104, USA}

\author{Ritesh Agarwal}
\affiliation{Department of Materials Science and Engineering, University of Pennsylvania, Philadelphia, Pennsylvania 19104, USA}

\author{Andrew M. Rappe}
\affiliation{Department of Chemistry, University of Pennsylvania, Philadelphia, Pennsylvania 19104, USA}
\email{rappe@sas.upenn.edu}

\date{\today}

\begin{abstract}
Spatially structured light carries independent phase and envelope variations in addition to its polarization and frequency. We develop a first-principles theory of second-order dc photocurrents (especially shift and injection) that resolves how these optical degrees of freedom drive distinct dc responses in crystals. The second-order dc response couples two optical field components at the same frequency but with generally different wave vectors. Expanding the response to first order in these wave vectors identifies two rank-four material tensors: $\xi$, which couples to the optical phase gradient and contains the plane-wave photon-drag limit, and $\zeta$, which couples to the intensity gradient and describes a spatially varying dc current. Wannier-interpolated calculations for monolayer GeS show symmetry-allowed finite-wave-vector responses that survive at a centrosymmetric structure where conventional bulk photovoltaic effect vanishes. For a Gaussian beam with a uniform phase gradient, the phase-gradient contribution drives current in the same direction on both sides of the beam center, whereas the intensity-gradient contribution drives currents in opposite directions, showing that distinct photocurrent patterns can be selected through the wavefront and envelope of light. A multipolar analysis further resolves the optical corrections into electric-quadrupole, orbital magnetic-dipole, and band-dispersion contributions. Together, these results show how the spatial structure of light can serve as an independent control variable for nonlinear photocurrent generation and provide a first-principles route to predict this response in real materials.
\end{abstract}

\maketitle

\blue{\textit{Introduction}.}---
The bulk photovoltaic effect (BPVE) converts light into a dc current without a junction or an applied bias~\cite{vonBaltzKraut1981,BelinicherSturman1980,AversaSipe1995,SipeShkrebtii2000,YoungRappe2012,TanEtAl2016,SpanierEtAl2016,DaiRappe2023}. In its conventional electric-dipole approximation, the optical field is treated as spatially uniform and the second-order response vanishes in centrosymmetric crystals~\cite{vonBaltzKraut1981,AversaSipe1995,SipeShkrebtii2000}. Finite photon momentum produces additional photocurrents, including photon-drag shift and injection responses~\cite{Shi2021,GassnerMele2023,XieNagaosa2025,Gassner2026}, while spatially varying illumination can drive local currents that disappear for a single plane wave~\cite{Ji2019}. These effects show that a crystal can respond not only to the polarization and frequency of light, but also to its spatial structure.

\begin{figure}[t]
\centering
\includegraphics[width=0.8\columnwidth]{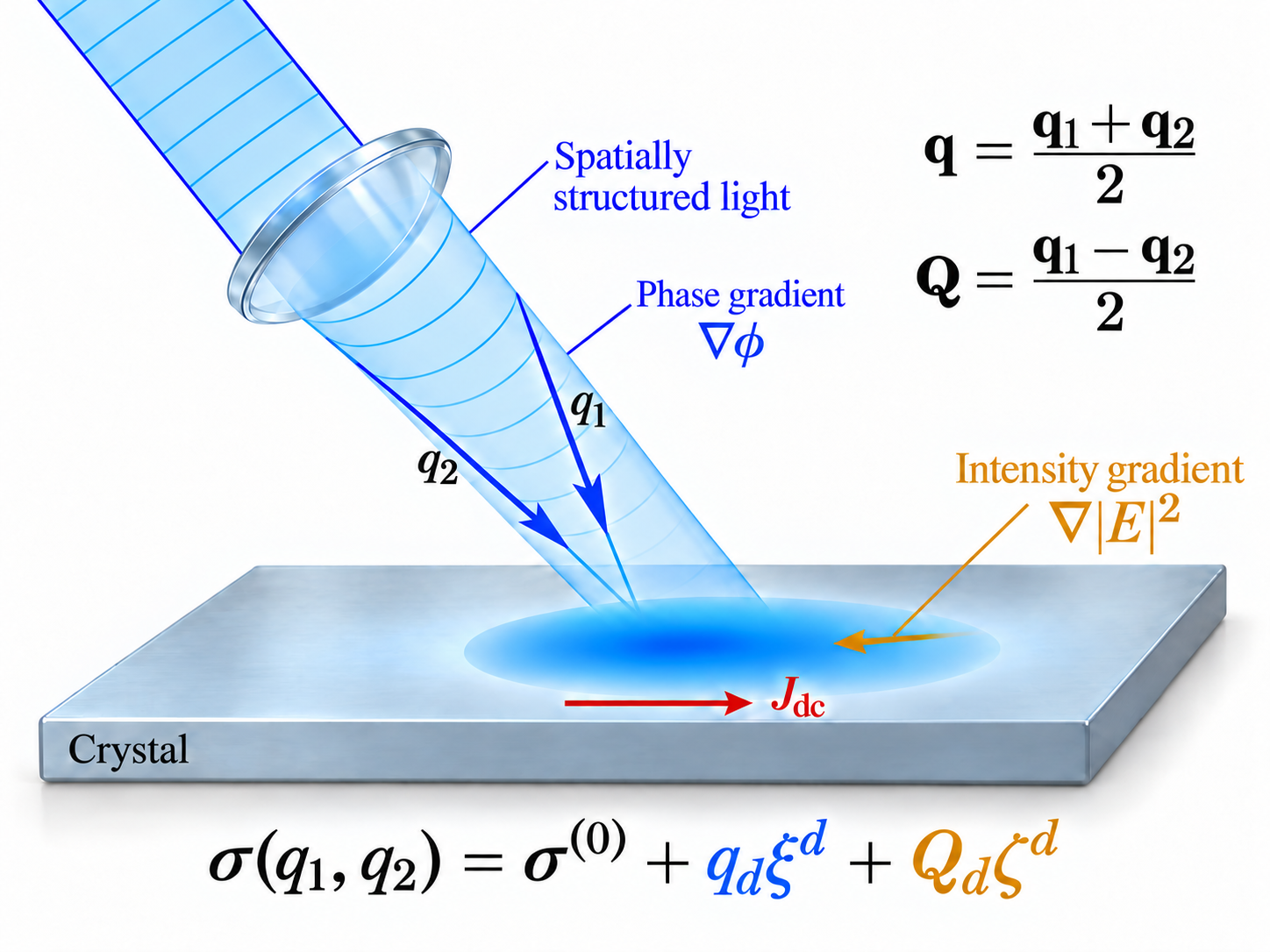}
\caption{Schmatic illustration of first-order finite-wave-vector photocurrents. Distinct optical Fourier components with wave vectors $\mathbf q_1$ and $\mathbf q_2$ define the average wave vector $\mathbf q=(\mathbf q_1+\mathbf q_2)/2$ and the dc output wave vector $\mathbf Q=(\mathbf q_1-\mathbf q_2)/2$. For a slowly varying field with fixed polarization, the $\xi$ and $\zeta$ responses couple to the optical phase gradient and intensity gradient, respectively.}
\label{fig:unified}
\end{figure}

A structured field contains more information than the momentum of one plane wave: its phase and envelope vary independently in space. For monochromatic light, the second-order dc response involves products of two optical field components,
\begin{equation}
E_b(\mathbf r)E_c^*(\mathbf r)= \int d\mathbf q_1d\mathbf q_2\,e^{i(\mathbf q_1-\mathbf q_2)\cdot\mathbf r}
E_b(\mathbf q_1)E_c^*(\mathbf q_2).
\label{eq:field_product_intro}
\end{equation}
The two components have the same frequency, so their time dependence cancels, but their wave vectors need not be equal. The average of the two wave vectors describes how the optical phase changes in space, while their difference describes the spatial modulation produced by interference between the two components (See Supplemental Material for detailed derivation). For a field with fixed polarization, these two combinations lead to photocurrent contributions associated with phase and intensity gradients, respectively as shown in Fig.~\ref{fig:unified}.
%Their difference therefore determines how the resulting dc response varies in space. 
It is noteworthy that whether these optical degrees of freedom can be separated into distinct material photocurrent responses and predicted from the electronic structure of a real crystal. 
%Their average controls the phase carried through the optical transition, whereas their difference sets the finite wave vector of the dc output. The latter is not determined by the plane-wave photon-drag limit because changing the output momentum also changes the measured-current vertex and its propagators.

Here we show that the average and difference momenta define two first-order material tensors, $\xi$ and $\zeta$, which couple directly to phase and intensity gradients of the light, respectively. We derive resonant shift- and injection-current kernels for both responses and evaluate their first-order wave-vector dependence using a first-principles-derived Wannier Hamiltonian. Our calculations for monolayer GeS show that both responses remain finite in the centrosymmetric phase, where the conventional BPVE vanishes. For a Gaussian beam with a uniform phase gradient, the phase-gradient contribution produces a current profile that is symmetric about the beam center, whereas the intensity-gradient contribution produces an antisymmetric profile. These distinct current patterns illustrate how the spatial phase and intensity profiles of light can control photocurrent generation. These findings pave the way for spatially resolved photodetection and optoelectronic control through the phase and intensity profiles of light.

\blue{\textit{General photocurrent response to spatially varying light}.}---
We consider the second-order dc current induced by monochromatic light whose amplitude and phase may vary in space. Building on Eq.~\eqref{eq:field_product_intro}, the current is written as~\cite{AversaSipe1995,SipeShkrebtii2000,Parker2019,GassnerMele2023}
\begin{equation}
J_a(\mathbf r)= \int d\mathbf q_1d\mathbf q_2\, e^{i(\mathbf q_1-\mathbf q_2)\cdot\mathbf r} \sigma^{abc}(\mathbf q_1,\mathbf q_2) E_b(\mathbf q_1)E_c^*(\mathbf q_2).
\label{eq:finite_q_response}
\end{equation}
Here $\sigma^{abc}$ describes the material response to a pair of optical Fourier components; the explicit resonant response kernels are given in the Supplemental Material. The integral includes all such pairs within the incident field. We focus on resonant interband processes: shift current arises from the displacement of charge during an optical transition, whereas injection current arises from the current carried by the photoexcited carrier population. To identify the leading effects of spatial variation, we Taylor-expand the response about $\mathbf q_1=\mathbf q_2=0$. We organize this expansion using the average and half-difference wave vectors,
$\mathbf q=(\mathbf q_1+\mathbf q_2)/2$ and
$\mathbf Q=(\mathbf q_1-\mathbf q_2)/2$:
\begin{equation}
\sigma_X^{abc}(\mathbf q+\mathbf Q,\mathbf q-\mathbf Q)
=
\sigma_{X,(0)}^{abc}
+q_d\xi_X^{abcd}
+Q_d\zeta_X^{abcd}
+O(q^2,Q^2,qQ),
\label{eq:qQ_expansion}
\end{equation}
where $X\in\{\mathrm{sh},\mathrm{inj}\}$ labels the shift and injection contributions. The zeroth-order term is the conventional BPVE response. The coefficients $\xi_X^{abcd}$ and $\zeta_X^{abcd}$ are the derivatives with respect to $q_d$ and $Q_d$, respectively, with both evaluated at $\mathbf q=\mathbf Q=0$. This first-order expansion describes the response in the long-wavelength limit, where higher-order wave-vector corrections can be neglected. All derivations are summarized in Supplemental Material.

The physical meaning of these coefficients follows by returning to real space. For a field with fixed polarization, we write $E_b(\mathbf r)=e_b\mathcal E(\mathbf r)$, where $e_b$ is a constant polarization vector and $\mathcal E(\mathbf r)=A(\mathbf r)e^{i\phi(\mathbf r)}=\int d\mathbf q_1\ e^{i\mathbf q_1\cdot\mathbf r}\mathcal E(\mathbf q_1)$. To express the wave-vector-dependent terms in Eq.~\eqref{eq:qQ_expansion} in real space, we use the relation between wave vectors and spatial derivatives of the Fourier factors. Since $-i\partial_d e^{i\mathbf q_1\cdot\mathbf r}=q_{1,d}e^{i\mathbf q_1\cdot\mathbf r}$ and $i\partial_d e^{-i\mathbf q_2\cdot\mathbf r}=q_{2,d}e^{-i\mathbf q_2\cdot\mathbf r}$, we obtain
\begin{align}
-i\mathcal E^*\partial_d\mathcal E
&=A^2\partial_d\phi-iA\partial_d A \nonumber\\
&=\int d\mathbf q_1d\mathbf q_2\ q_{1,d}e^{i(\mathbf q_1-\mathbf q_2)\cdot\mathbf r} \mathcal E(\mathbf q_1)\mathcal E^*(\mathbf q_2),
\label{eq:q1_realspace}
\\
i\mathcal E\partial_d\mathcal E^*
&=A^2\partial_d\phi+iA\partial_d A \nonumber\\
&=\int d\mathbf q_1d\mathbf q_2\ q_{2,d}e^{i(\mathbf q_1-\mathbf q_2)\cdot\mathbf r} \mathcal E(\mathbf q_1)\mathcal E^*(\mathbf q_2).
\label{eq:q2_realspace}
\end{align}
The term proportional to $q_d=(q_{1,d}+q_{2,d})/2$ in Eq.~\eqref{eq:qQ_expansion} is obtained by averaging these two expressions. The amplitude-gradient terms cancel, leaving $A^2\partial_d\phi$, which is proportional to the local intensity multiplied by the phase gradient. The term proportional to $Q_d=(q_{1,d}-q_{2,d})/2$ is obtained by taking half their difference. The phase-gradient terms then cancel, leaving $-iA\partial_d A=-(i/2)\partial_d A^2$, which depends only on the intensity gradient. Substituting the expansion into Eq.~\eqref{eq:finite_q_response} therefore gives
\begin{align}
J_{a,X}^{(\xi)}(\mathbf r)=\xi_X^{abcd}e_be_c^*A^2\partial_d\phi,
\label{eq:realspace_phase_intensity1}
\\
J_{a,X}^{(\zeta)}(\mathbf r)=-\frac{i}{2}\zeta_X^{abcd}e_be_c^*\partial_d A^2.
\label{eq:realspace_phase_intensity}
\end{align}
Here, we describe two distinct ways to generate photocurrents driven by spatial variations of the optical phase ($\xi_X$) and intensity ($\zeta_X$), respectively as illustrated in Fig.~\ref{fig:unified}. Their connection to photon drag becomes clear for a single plane wave, $\mathcal E(\mathbf r)=A_0e^{i\mathbf q_0\cdot\mathbf r}$. In this limit, the intensity is uniform and $\boldsymbol\nabla\phi=\mathbf q_0$, so the $\zeta_X$ contribution vanishes while the $\xi_X$ contribution is proportional to the optical wave vector, recovering the plane-wave photon-drag response. For spatially varying illumination, both contributions can coexist. 
%Together, they describe the first-order spatial dispersion of the second-order conductivity: the material response depends on the optical wave vectors and therefore on spatial gradients of the field, beyond its local value. 
The full microscopic derivations of $\xi_X$ and $\zeta_X$ for both shift and injection currents are provided in the Supplemental Material. These gradient-dependent currents are allowed even in centrosymmetric crystals. Under inversion, the current changes sign while the local quadratic field product remains unchanged, forbidding the conventional BPVE. The phase and intensity gradients also change sign, however, allowing the responses described by $\xi_X$ and $\zeta_X$, subject to the remaining crystal symmetries. We evaluate these coefficients using a first-principles-derived Wannier Hamiltonian and use monolayer GeS to examine how crystal symmetry and the spatial profile of light determine the resulting photocurrents.

\begin{figure}[t]
\centering
\includegraphics[width=0.9\columnwidth]{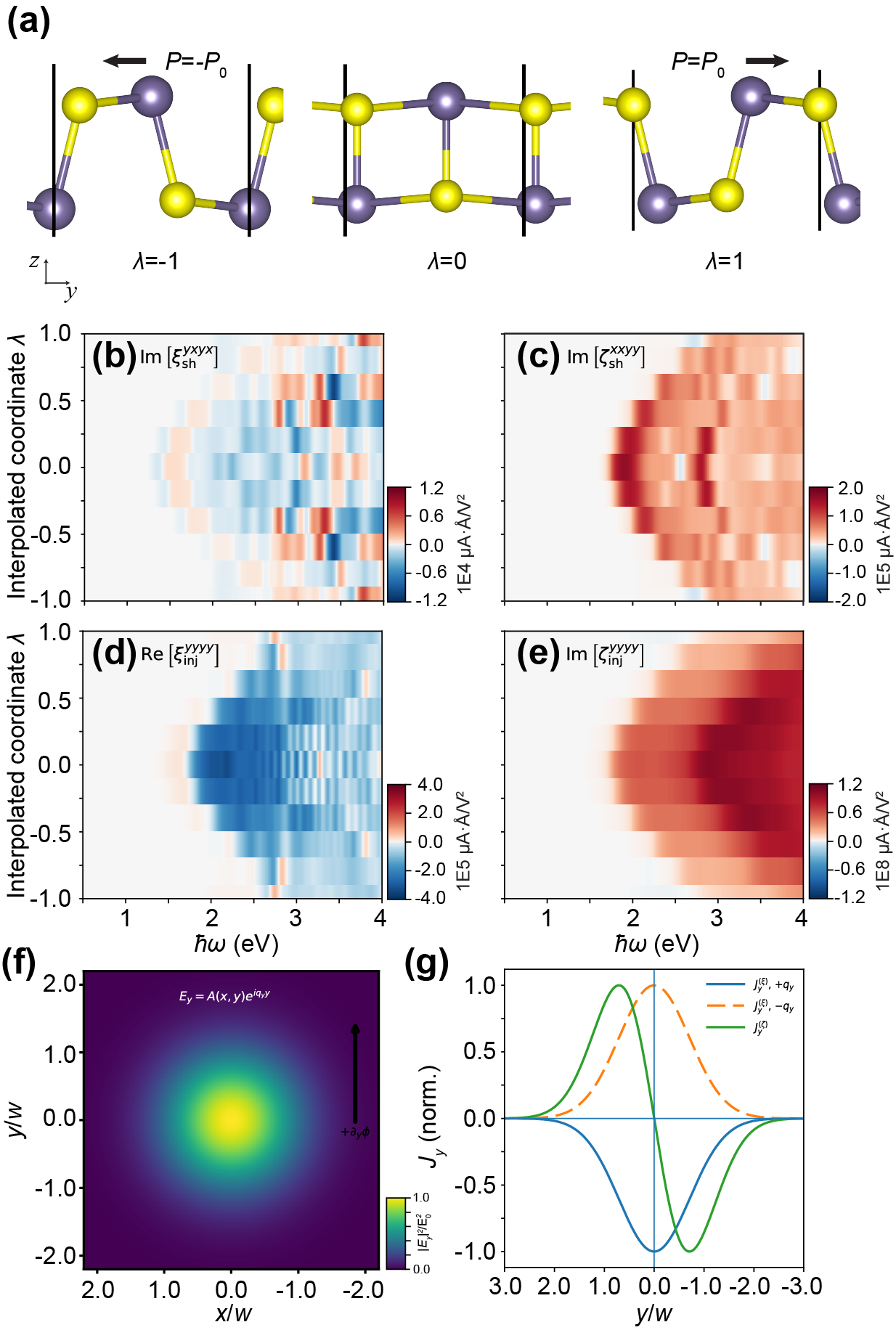}
\caption{First-principles finite-wave-vector photocurrent response of monolayer GeS. (a) Structural interpolation between inversion-related polar states at $\lambda=\pm1$ through a centrosymmetric reference at $\lambda=0$. (b)--(e) Photon-energy and structural-coordinate dependence of representative symmetry-allowed components: (b) $\operatorname{Im}\xi_{\rm sh}^{yyxx}$, (c) $\operatorname{Im}\zeta_{\rm sh}^{xxyy}$, (d) $\operatorname{Re}\xi_{\rm inj}^{yyyy}$, and (e) $\operatorname{Im}\zeta_{\rm inj}^{yyyy}$. The plotted coefficients use the effective-three-dimensional normalization $\sigma_{3D}=L \sigma_{2D}$. (f) Gaussian structured field profile $E_y(x,y)=E_0e^{-(x^2+y^2)/(2w^2)}e^{iq_yy}$, where $w$ is the $1/e$ intensity radius. (g) Normalized current profiles at $\lambda=0$ and $\hbar\omega=1.98$~eV along $x=0$.}
\label{fig:GeS}
\end{figure}

\blue{\textit{First-principles benchmark and illumination control in GeS}.}---
We apply the theory to monolayer GeS, a polar group-IV monochalcogenide with large conventional BPVE responses~\cite{Cook2017,Rangel2017,Panday2019,FeiTanRappe2020}. To examine whether the gradient-dependent photocurrents require a polar structure, we interpolate between two inversion-related polar configurations at $\lambda=\pm1$ through a centrosymmetric reference at $\lambda=0$, as shown in Fig.~\ref{fig:GeS}(a). The conventional BPVE vanishes at this reference structure, allowing us to isolate the responses driven by spatial variations of the optical field. Figures~\ref{fig:GeS}(b)-(e) show representative components allowed by the crystal point-group symmetries: $\operatorname{Im}\xi_{\rm sh}^{yyxx}$, $\operatorname{Im}\zeta_{\rm sh}^{xxyy}$, $\operatorname{Re}\xi_{\rm inj}^{yyyy}$, and $\operatorname{Im}\zeta_{\rm inj}^{yyyy}$. We report effective three-dimensional coefficients in units of $\mu\mathrm{A}\,\mathrm{\AA}/\mathrm{V}^{2}$, so that multiplication by $q_d$ for $\xi$ or $Q_d$ for $\zeta$ gives a conductivity scale in $\mu\mathrm{A}/\mathrm{V}^{2}$. For an order-of-magnitude estimate, we take $q_0=10^{-3}\,\mathrm{\AA}^{-1}$, corresponding to the vacuum photon wave vector $q_0=\omega/c$ at $\hbar\omega\simeq2$~eV. and $Q_0=10^{-4}\,\mathrm{\AA}^{-1}$, corresponding to a spatial modulation period $L=2\pi/Q_0\simeq6.3\,\mu\mathrm{m}$. With these values, the calculated responses correspond to conductivity scales of approximately $10^{1}$--$10^{4}\,\mu\mathrm{A}/\mathrm{V}^{2}$, compared with conventional shift conductivities of order $10^{2}\,\mu\mathrm{A}/\mathrm{V}^{2}$ reported for monolayer group-IV monochalcogenides~\cite{Rangel2017}. 
%This comparison is intended only as an order-of-magnitude reference, particularly for the injection responses, which depend on the assumed relaxation time. 
The coefficients are even under $\lambda\rightarrow-\lambda$ and remain finite at $\lambda=0$, where the conventional BPVE is forbidden. Thus, unlike the conventional BPVE, we confirm that these photocurrents do not require broken inversion symmetry. Reversing the polar distortion leaves the first-order gradient coefficients unchanged, while the spatial structure of the illumination provides another means of controlling the current.

We next examine how the phase and intensity gradients produce different current patterns in the same material. At the centrosymmetric structure, both $\operatorname{Re}\xi_{\rm inj}^{yyyy}$ and $\operatorname{Im}\zeta_{\rm inj}^{yyyy}$ are finite over a visible light range. These components describe a $y$-directed injection current driven by $y$-polarized light with spatial variations along $y$, allowing the two responses to be compared in the same geometry. We consider a beam whose dominant electric-field component in the sample plane is
\begin{equation}
E_y(x,y)=E_0e^{-(x^2+y^2)/(2w^2)}e^{iq_y y},
\label{eq:ges_beam}
\end{equation}
as illustrated in Fig.~\ref{fig:GeS}(f). Here $w$ is the $1/e$ intensity radius and $q_y$ is the uniform in-plane phase gradient. To evaluate the two contributions, we focus on the central line $x=0$, where the transverse intensity gradient vanishes ($\left.\partial_x |E_y(x,y)|^2 \right|_{x=0}=0$). Along this line, the intensity is $A^2=E_0^2e^{-y^2/w^2}$ and the phase is $\phi=q_y y$. The phase therefore has a constant gradient, $\partial_y\phi=q_y$, whereas the intensity gradient is $\partial_y A^2=-(2y/w^2)A^2$. Substituting these expressions into Eqs.~\eqref{eq:realspace_phase_intensity1} and \eqref{eq:realspace_phase_intensity}
and taking the real parts gives the $yyyy$ contributions:
\begin{align}
\frac{J_{y,\rm inj}^{(\xi)}}{E_0^2}
&=q_y\operatorname{Re}\xi_{\rm inj}^{yyyy}e^{-u^2},
\\
\frac{J_{y,\rm inj}^{(\zeta)}}{E_0^2}
&=-\frac{\operatorname{Im}\zeta_{\rm inj}^{yyyy}}{w}\,u e^{-u^2},
\label{eq:ges_profile}
\end{align}
where $u=y/w$. where $u=y/w$. The phase-gradient contribution thus follows the Gaussian intensity, with the same current direction on both sides of the beam center. In contrast, the intensity-gradient contribution vanishes at the center and changes sign across it, where the intensity changes from increasing to decreasing. Reversing $q_y$ at fixed intensity profile reverses only the phase-gradient contribution. We plot these even and odd spatial profiles at $\lambda=0$ and $\hbar\omega=1.98$~eV in Figure~\ref{fig:GeS}(g). Each curve is normalized by its own maximum magnitude to highlight its spatial shape; the absolute magnitudes follow from Eq.~\eqref{eq:ges_profile} and the calculated coefficients in Figs.~\ref{fig:GeS}(d),(e). The different spatial profiles and responses to phase-gradient reversal provide a way to distinguish the two contributions.

\begin{figure}[t]
    \centering
    \includegraphics[width=\linewidth]{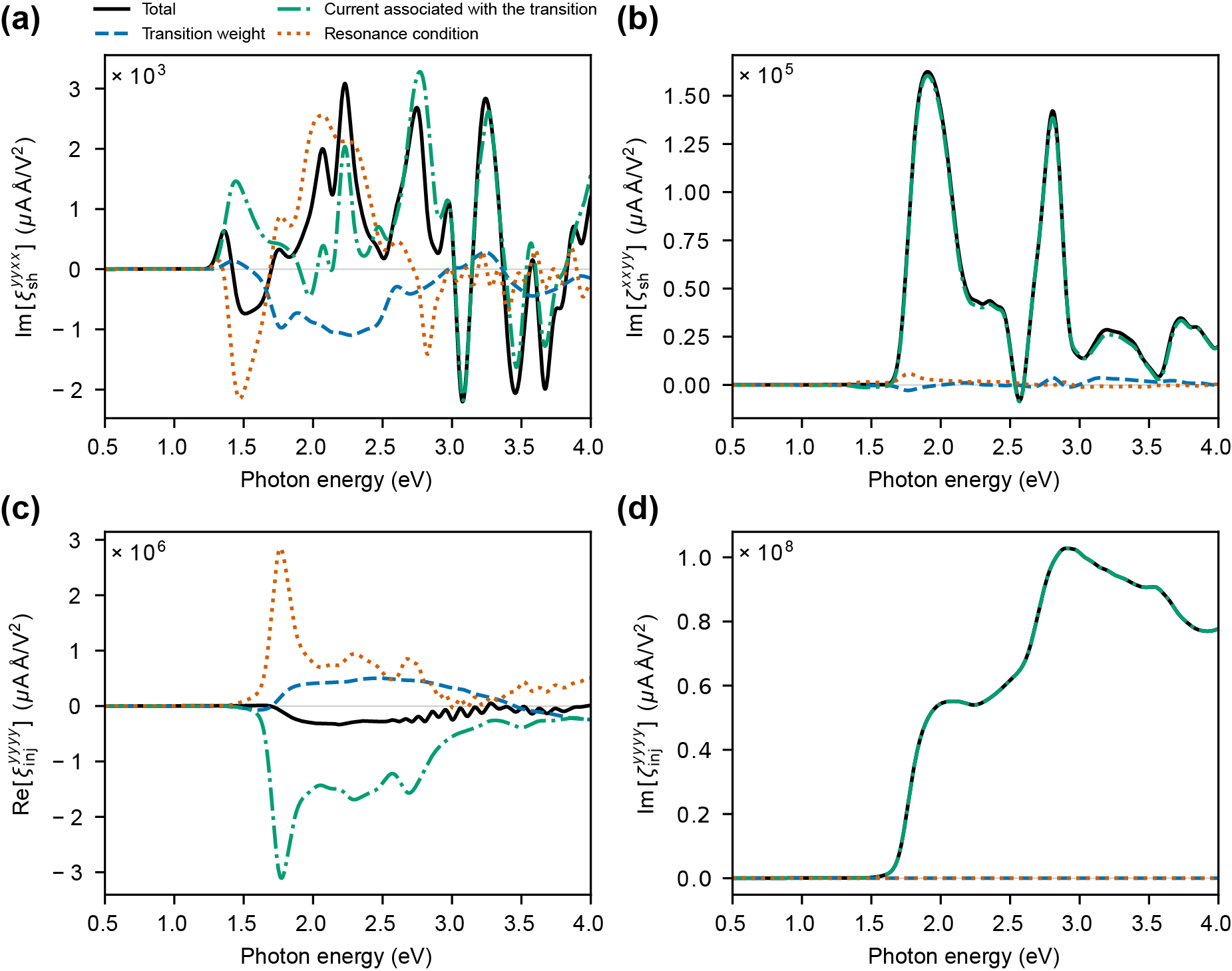}
    \caption{Decomposition of finite-wave-vector photocurrent responses in centrosymmetric GeS ($\lambda=0$). (a) $\operatorname{Im}\xi_{\rm sh}^{yyxx}$, (b) $\operatorname{Im}\zeta_{\rm sh}^{xxyy}$, (c) $\operatorname{Re}\xi_{\rm inj}^{yyyy}$, and (d) $\operatorname{Im}\zeta_{\rm inj}^{yyyy}$. Black solid curves show the total response, while blue dashed, green dash-dotted, and orange dotted curves show the transition-weight, current-response, and resonance contributions, respectively.}
    \label{fig:ges_decomposition}
\end{figure}

\blue{\textit{Discussion}.}---To understand the origin of the large finite-wave-vector responses in GeS, we decompose the derived response kernels into contributions arising from the wave-vector dependence of their constituent factors. For a fixed linear polarization along $b$, the conventional BPVE can be written schematically as~\cite{SipeShkrebtii2000}
\begin{align}
J_{\rm sh}^{a}&\propto
\sum_{nm}\int_{\rm BZ}d\mathbf k\,
\mathcal W_{mn}^{b}R_{mn}^{a;b}
\delta(\omega_{mn}-\omega),
\\
\dot J_{\rm inj}^{a}
&\propto
\sum_{nm}\int_{\rm BZ}d\mathbf k\,
\mathcal W_{mn}^{b}\Delta v_{mn}^{a}
\delta(\omega_{mn}-\omega),
\label{eq:discussion_local_structure}
\end{align}
where $\mathcal W_{mn}^{b}=(f_n-f_m)|r_{mn}^{b}|^2$ is the occupation-weighted transition strength, $R_{mn}^{a;b}$ is the shift vector, and $\Delta v_{mn}^{a}=v_m^a-v_n^a$. At finite optical wave vectors, the transition amplitudes, transition energies, and current-response factors acquire additional wave-vector dependence. To illustrate how the resulting terms arise, consider a resonant contribution written schematically as $\mathcal W(s)\mathcal T(s)\delta[\Omega(s)-\omega]$, where $s$ is a component of $\mathbf q$ or $\mathbf Q$ and $\mathcal T$ denotes the associated current-response factor. Its first-order correction is
\begin{align}
\partial_s\!\left[\mathcal W\mathcal T
\delta(\Omega-\omega)\right]
&={}
(\partial_s\mathcal W)\mathcal T
\delta(\Omega-\omega)
\nonumber\\
&+\mathcal W(\partial_s\mathcal T)
\delta(\Omega-\omega)
\nonumber\\
&+\mathcal W\mathcal T(\partial_s\Omega)
\delta'(\Omega-\omega),
\label{eq:discussion_gradient_structure}
\end{align}
with all quantities evaluated at the expansion point. The three terms describe changes in the transition weight, the current associated with the transition, and the resonance condition, respectively. We apply this decomposition to the four tensor components exhibiting large responses in Figs.~\ref{fig:GeS}(b)--(e): $\operatorname{Im}\xi_{\rm sh}^{yyxx}$, $\operatorname{Im}\zeta_{\rm sh}^{xxyy}$, $\operatorname{Re}\xi_{\rm inj}^{yyyy}$, and $\operatorname{Im}\zeta_{\rm inj}^{yyyy}$. For $\operatorname{Im}\xi_{\rm sh}^{yyxx}$ and $\operatorname{Re}\xi_{\rm inj}^{yyyy}$ [Fig.~\ref{fig:ges_decomposition}(a),(c)], the total spectrum cannot be inferred from any single contribution. Thus, the magnitude and sign of these $\xi$ components reflect the balance among several wave-vector-dependent factors. In contrast, the current-response contribution closely follows the total for both $\operatorname{Im}\zeta_{\rm sh}^{xxyy}$ and $\operatorname{Im}\zeta_{\rm inj}^{yyyy}$ [Figs.~\ref{fig:ges_decomposition}(b),(d)]. It accounts for the prominent peaks near $1.9$ and $2.8$~eV in the shift response and the broad positive injection response, whereas the transition-weight and resonance contributions are much smaller on the corresponding scales. For these components, the large response therefore arises primarily from the remaining current-response kernel rather than from changes in the optical transition weight or resonance condition. These results highlight the wave-vector sensitivity of the current generated by optical transitions as an important ingredient in understanding large spatially dispersive photocurrents.

%The distinction between phase- and intensity-gradient responses is complementary to the microscopic classification of light-matter interactions. The first-order corrections to the optical transition matrix elements include electric-quadrupole and orbital magnetic-dipole terms, which interfere with the leading electric-dipole coupling, together with contributions from electronic band dispersion~\cite{MalashevichSouza2010,PozoSouza2023,Urru2025}. These optical corrections contribute to both $\xi_X$ and $\zeta_X$; the two coefficients therefore cannot be identified separately with electric-quadrupole and magnetic-dipole effects. Instead, $\xi_X$ and $\zeta_X$ distinguish the spatial features of light that drive the current, while the multipolar decomposition identifies microscopic contributions to the response. The complete coefficients also include wave-vector derivatives of the other factors in the response kernels. The detailed decomposition is presented in the Supplemental Material.

\blue{\textit{Conclusion}.}---
We have developed a unified first-principles framework for shift and injection photocurrents driven by the spatial structure of light. Expanding in the average and difference of the two optical wave vectors yields distinct first-order tensors that couple to phase and intensity gradients of structured field. This place plane-wave photon drag and the finite-output spatial-gradient response within a single spatially dispersive framework. In GeS, we confirmed both responses remain finite at the centrosymmetric structure. Interestingly, the $yyyy$ injection components generate distinct even and odd spatial current profiles, and reversing the phase gradient at fixed intensity reverses only the phase-driven contribution. Decomposing the response kernels further reveals that the magnitude and sign of the calculated coefficients depend on the balance among optical-transition, resonance, and current-response contributions. Together, these results connect the microscopic origins of spatially dispersive photocurrents to their control through the phase and intensity profiles of light.

\blue{\emph{Acknowledgment}}.--- This work was supported by the U.S. Department of Energy, Office of Science, Basic Energy Sciences, under Award No. DE-SC0024942. Computational support was provided by the National Energy Research Scientific Computing Center (NERSC), a U.S. Department of Energy, Office of Science User Facility located at Lawrence Berkeley National Laboratory, operated under Contract No. DE-AC02-05CH11231.

\blue{\emph{Data Availability}}.--- The data supporting the findings of this study will be deposited in the Dryad Digital Repository upon publication~\cite{Kim2026Dryad}.

\bibliographystyle{apsrev4-2}
\bibliography{references_unified_additions_v2}

\begin{thebibliography}{99}

\bibitem[S1]{SM:Shi2021}
L.-k. Shi, D. Zhang, K. Chang, and J. C. W. Song,
\href{https://doi.org/10.1103/PhysRevLett.126.197402}{Phys. Rev. Lett. \textbf{126}, 197402 (2021)}.

\bibitem[S2]{SM:GassnerMele2023}
S. Gassner and E. J. Mele,
\href{https://doi.org/10.1103/PhysRevB.108.085403}{Phys. Rev. B \textbf{108}, 085403 (2023)}.

\bibitem[S3]{SM:XieNagaosa2025}
Y.-M. Xie and N. Nagaosa,
\href{https://doi.org/10.1073/pnas.2424294122}{Proc. Natl. Acad. Sci. U.S.A. \textbf{122}, e2424294122 (2025)}.

\bibitem[S4]{SM:AversaSipe1995}
C. Aversa and J. E. Sipe,
\href{https://doi.org/10.1103/PhysRevB.52.14636}{Phys. Rev. B \textbf{52}, 14636 (1995)}.

\bibitem[S5]{SM:SipeShkrebtii2000}
J. E. Sipe and A. I. Shkrebtii,
\href{https://doi.org/10.1103/PhysRevB.61.5337}{Phys. Rev. B \textbf{61}, 5337 (2000)}.

\bibitem[S6]{SM:Parker2019}
D. E. Parker, T. Morimoto, J. Orenstein, and J. E. Moore,
\href{https://doi.org/10.1103/PhysRevB.99.045121}{Phys. Rev. B \textbf{99}, 045121 (2019)}.

\bibitem[S7]{SM:Urru2025}
A. Urru, I. Souza, \'O. Pozo Oca\~na, S. S. Tsirkin, and D. Vanderbilt,
\href{https://doi.org/10.1103/PhysRevB.112.045201}{Phys. Rev. B \textbf{112}, 045201 (2025)}.

\end{thebibliography}

% ======================================================================
% SUPPLEMENTAL MATERIAL (one-column format, independent S-numbering)
% ======================================================================
\clearpage
\onecolumngrid
\setcounter{secnumdepth}{5}
\setcounter{section}{0}
\setcounter{equation}{0}
\setcounter{figure}{0}
\setcounter{table}{0}
\renewcommand{\thesection}{S\arabic{section}}
\renewcommand{\theequation}{S\arabic{equation}}
\renewcommand{\thefigure}{S\arabic{figure}}
\renewcommand{\thetable}{S\arabic{table}}
% The supplement was originally a single-column article. In single-column
% REVTeX, widetext should be a no-op, rather than switching column modes.
\renewenvironment{widetext}{}{}
\setlength{\parindent}{0pt}
\setlength{\parskip}{4pt}
% The supplement uses independently numbered references [S1], [S2], ...
% with keys SM:... to prevent clashes with manuscript BibTeX entries.
\bibpunct{[}{]}{,}{n}{}{,}
\begin{center}
{\Large\bfseries Supplemental Material for\\[4pt]
``Unified First-Principles Theory of Finite-Wave-Vector Photocurrents in Crystals''\par}
\vspace{1em}
{\normalsize Bumseop Kim, Sayed Ali Akbar Ghorashi, Yicong Chen,\\
Ritesh Agarwal, and Andrew M. Rappe\par}
\end{center}
\vspace{0.6em}

\section{Real-space interpretation of the average and difference optical wave vectors}

For monochromatic illumination, the positive-frequency electric field can be written as
\begin{equation}
E_b^{(+)}(\mathbf r,t)
=
E_b(\mathbf r)e^{-\ii\omega t}.
\end{equation}
The quadratic field product entering the dc response is therefore
\begin{equation}
E_b^{(+)}(\mathbf r,t)
\left[E_c^{(+)}(\mathbf r,t)\right]^*
=
E_b(\mathbf r)E_c^*(\mathbf r).
\end{equation}
The optical time dependence cancels because the two field components have the same frequency; their wave vectors, however, need not be equal.

Using the Fourier convention
\begin{equation}
E_b(\mathbf r)
=
\int d\mathbf q_1\,
e^{\ii\mathbf q_1\cdot\mathbf r}E_b(\mathbf q_1),
\label{eq:FourierConventionS}
\end{equation}
the quadratic field product becomes
\begin{equation}
E_b(\mathbf r)E_c^*(\mathbf r)
=
\int d\mathbf q_1d\mathbf q_2\,
e^{\ii(\mathbf q_1-\mathbf q_2)\cdot\mathbf r}
E_b(\mathbf q_1)E_c^*(\mathbf q_2).
\label{eq:fieldProductS}
\end{equation}
We define the average and half-difference wave vectors as
\begin{equation}
\mathbf q=\frac{\mathbf q_1+\mathbf q_2}{2},
\qquad
\mathbf Q=\frac{\mathbf q_1-\mathbf q_2}{2}.
\label{eq:qQdefinitionsS}
\end{equation}
Thus, the spatial modulation associated with each pair of optical components is described by
$e^{\ii(\mathbf q_1-\mathbf q_2)\cdot\mathbf r}
=e^{2\ii\mathbf Q\cdot\mathbf r}$.

To identify the real-space meaning of the first-order terms, consider a field with spatially uniform polarization,
\begin{equation}
E_b(\mathbf r)=e_b\mathcal E(\mathbf r),
\qquad
\mathcal E(\mathbf r)=A(\mathbf r)e^{\ii\phi(\mathbf r)},
\label{eq:scalarfieldS}
\end{equation}
where $\mathbf e^\dagger\mathbf e=1$, $A(\mathbf r)\geq0$ is the field amplitude, and $\phi(\mathbf r)$ is the phase. The intensity is proportional to $A^2(\mathbf r)$. The scalar field obeys the same Fourier convention as Eq.~\eqref{eq:FourierConventionS}, giving
\begin{equation}
A^2(\mathbf r)
=
\int d\mathbf q_1d\mathbf q_2\,
e^{\ii(\mathbf q_1-\mathbf q_2)\cdot\mathbf r}
\mathcal E(\mathbf q_1)\mathcal E^*(\mathbf q_2).
\label{eq:intensityFourierS}
\end{equation}
Writing $\partial_d=\partial/\partial r_d$, differentiation of the two Fourier factors yields
\begin{align}
&\int d\mathbf q_1d\mathbf q_2\,
e^{\ii(\mathbf q_1-\mathbf q_2)\cdot\mathbf r}
q_{1,d}\mathcal E(\mathbf q_1)\mathcal E^*(\mathbf q_2)
\nonumber\\
&\qquad=-\ii(\partial_d\mathcal E)\mathcal E^*,
\label{eq:q1FourierS}\\
&\int d\mathbf q_1d\mathbf q_2\,
e^{\ii(\mathbf q_1-\mathbf q_2)\cdot\mathbf r}
q_{2,d}\mathcal E(\mathbf q_1)\mathcal E^*(\mathbf q_2)
\nonumber\\
&\qquad=\ii\mathcal E(\partial_d\mathcal E^*).
\label{eq:q2FourierS}
\end{align}

The average wave vector therefore gives
\begin{align}
&\int d\mathbf q_1d\mathbf q_2\,
e^{\ii(\mathbf q_1-\mathbf q_2)\cdot\mathbf r}
q_d\mathcal E(\mathbf q_1)\mathcal E^*(\mathbf q_2)
\nonumber\\
&\qquad=
-\frac{\ii}{2}
\left[
\mathcal E^*\partial_d\mathcal E
-\mathcal E\partial_d\mathcal E^*
\right].
\label{eq:qaverageIntermediateS}
\end{align}
Substituting Eq.~\eqref{eq:scalarfieldS}, we obtain
\begin{align}
\mathcal E^*\partial_d\mathcal E
&=A\partial_dA+\ii A^2\partial_d\phi,
\\
\mathcal E\partial_d\mathcal E^*
&=A\partial_dA-\ii A^2\partial_d\phi.
\end{align}
The amplitude-gradient terms cancel in their difference, so
\begin{equation}
\int d\mathbf q_1d\mathbf q_2\,
e^{\ii(\mathbf q_1-\mathbf q_2)\cdot\mathbf r}
q_d\mathcal E(\mathbf q_1)\mathcal E^*(\mathbf q_2)
=
A^2\partial_d\phi.
\label{eq:phasegradientS}
\end{equation}
The average wave vector thus generates the intensity-weighted phase gradient.

By contrast, the half-difference wave vector gives
\begin{align}
&\int d\mathbf q_1d\mathbf q_2\,
e^{\ii(\mathbf q_1-\mathbf q_2)\cdot\mathbf r}
Q_d\mathcal E(\mathbf q_1)\mathcal E^*(\mathbf q_2)
\nonumber\\
&\qquad=
-\frac{\ii}{2}
\left[
\mathcal E^*\partial_d\mathcal E
+\mathcal E\partial_d\mathcal E^*
\right]
\nonumber\\
&\qquad=
-\frac{\ii}{2}\partial_dA^2.
\label{eq:intensitygradientS}
\end{align}
Here the phase-gradient terms cancel, leaving the intensity gradient.

Restoring the polarization factors, the two identities are
\begin{align}
&\int d\mathbf q_1d\mathbf q_2\,
e^{\ii(\mathbf q_1-\mathbf q_2)\cdot\mathbf r}
q_d E_b(\mathbf q_1)E_c^*(\mathbf q_2)
\nonumber\\
&\qquad=e_be_c^*A^2\partial_d\phi,
\\
&\int d\mathbf q_1d\mathbf q_2\,
e^{\ii(\mathbf q_1-\mathbf q_2)\cdot\mathbf r}
Q_d E_b(\mathbf q_1)E_c^*(\mathbf q_2)
\nonumber\\
&\qquad=-\frac{\ii}{2}e_be_c^*\partial_dA^2.
\label{eq:FourierIdentitiesS}
\end{align}
Consequently, the terms proportional to $q_d\xi_X^{abcd}$ and $Q_d\zeta_X^{abcd}$ in the response kernel yield
\begin{align}
J_{a,X}^{(\xi)}(\mathbf r)
&=
\xi_X^{abcd}e_be_c^*A^2\partial_d\phi,
\\
J_{a,X}^{(\zeta)}(\mathbf r)
&=
-\frac{\ii}{2}\zeta_X^{abcd}e_be_c^*\partial_dA^2,
\label{eq:realspaceGradientResponsesS}
\end{align}
where $X$ denotes the shift or injection contribution and repeated Cartesian indices are summed. These Fourier identities hold even when the amplitude and phase vary simultaneously in space. They are exact for a differentiable field with fixed polarization; the first-order approximation enters when the material response kernel is truncated at linear order in $\mathbf q$ and $\mathbf Q$.

\section{Finite-wave-vector response and resonant kernels}
\label{sec:density}

We consider the resonant interband photocurrent of a band insulator under monochromatic illumination. The second-order dc current is
\begin{equation}
J_a(\mathbf r)
=
\int d\mathbf q_1d\mathbf q_2\,
e^{\ii(\mathbf q_1-\mathbf q_2)\cdot\mathbf r}
\sigma^{abc}(\mathbf q_1,\mathbf q_2)
E_b(\mathbf q_1)E_c^*(\mathbf q_2),
\label{eq:finite_q_responseS}
\end{equation}
where the two optical components have the same frequency $\omega$ but may have different wave vectors. We use
$\mathbf q=(\mathbf q_1+\mathbf q_2)/2$ and
$\mathbf Q=(\mathbf q_1-\mathbf q_2)/2$, so the spatial dependence of each contribution is $e^{2\ii\mathbf Q\cdot\mathbf r}$. Throughout this section, $e>0$ is the elementary charge, the electron charge is $-e$, and
\begin{equation}
f_{\alpha\beta}=f_\alpha-f_\beta,
\qquad
\omega_{\alpha\beta}
=\frac{\varepsilon_\alpha-\varepsilon_\beta}{\hbar}.
\end{equation}
The notation $\sum_{nm\mathbf k}$ includes the Brillouin-zone integral
$\int_{\rm BZ}d^Dk/(2\pi)^D$.

\subsection{Nonvertical optical transitions}

A finite optical wave vector connects electronic states at different crystal momenta. For example, the positive-frequency field at $\mathbf q_1$ drives
\begin{equation}
|m,\mathbf k\rangle
\longrightarrow
|n,\mathbf k+\mathbf q_1\rangle.
\label{eq:nonvertical_transitionS}
\end{equation}
This is the nonvertical transition underlying the photon-drag response~\cite{SM:Shi2021,SM:GassnerMele2023,SM:XieNagaosa2025}. To keep the expressions compact, we write
\begin{equation}
n=(n,\mathbf k),\qquad
n_i=(n,\mathbf k+\mathbf q_i),\qquad
n_{12}=(n,\mathbf k+2\mathbf Q),
\label{eq:band_labelsS}
\end{equation}
and use the same convention for other band labels.

Both optical interactions and the output current are described using finite-momentum velocity matrix elements,
\begin{equation}
v^a_{n\mathbf k',m\mathbf k}
=
\langle\psi_{n\mathbf k'}|
\hat v^a(\mathbf k'-\mathbf k)
|\psi_{m\mathbf k}\rangle.
\label{eq:finite_momentum_vertexS}
\end{equation}
Here $\hat v^a(\mathbf p)$ carries momentum $\mathbf p$, and
$\hat v^a(\mathbf 0)$ is the ordinary velocity operator. Thus,
$v^b_{n_1m}$ describes the optical transition at $\mathbf q_1$,
whereas $v^a_{l_2n_1}$ and $v^a_{ml_{12}}$ close the response with momentum transfer $-2\mathbf Q$.
The field-dependent current also contributes a contact matrix element
$w^{ac}_{mn_1}(-2\mathbf Q,-\mathbf q_2)$.
Its normalization is defined by
\begin{equation}
\hat J_a(2\mathbf Q;A)
=
-e\hat v^a(-2\mathbf Q)
-\frac{e^2}{\hbar}
\sum_c\int d\mathbf p\,
A_c(\mathbf p)\hat w^{ac}(-2\mathbf Q,\mathbf p)
+O(A^2),
\label{eq:current_vertexS}
\end{equation}
where $A_c$ denotes the vector potential.
The explicit velocity and contact operators used in the calculation are specified in the vertex construction below.

\subsection{Resonant contribution}

The resonant response follows from the optical poles of second-order perturbation theory~\cite{SM:GassnerMele2023,SM:AversaSipe1995,SM:SipeShkrebtii2000,SM:Parker2019}. In the temporal gauge, where the scalar potential vanishes
($\phi=0$), $A_b^{(+)}=E_b/(i\omega)$, and the first-order
density matrix associated with Eq.~\eqref{eq:nonvertical_transitionS} is
\begin{equation}
\rho_{n_1m}^{(1,+)}
=
\frac{eA_b^{(+)}f_{n_1m}v^b_{n_1m}}
{\hbar(\omega_{n_1m}-\omega-\ii\eta)}.
\label{eq:first_order_resonantS}
\end{equation}
Using
\begin{equation}
\frac{1}{x-\ii0^+}
=
\mathcal P\frac{1}{x}
+\ii\pi\delta(x),
\end{equation}
we retain the optical delta-function contribution,
\begin{equation}
\rho_{n_1m}^{(1,+)}\big|_{\rm res}
=
\frac{\ii\pi e}{\hbar}
A_b^{(+)}f_{n_1m}v^b_{n_1m}
\delta(\omega_{n_1m}-\omega).
\label{eq:optical_poleS}
\end{equation}
The second interaction, at wave vector $-\mathbf q_2$ and frequency $-\omega$, can act on either side of this density matrix. The two resulting paths contain output denominators $\omega_{n_1l_2}-2\ii\eta$ and $\omega_{l_{12}m}-2\ii\eta$, with opposite signs from the commutator. The factor $2\eta$ in the dc output denominators follows from the adiabatic switching factor $e^{\eta t}$ associated with each optical interaction. The two resulting pathways give the three-vertex contributions, while the field-dependent part of the current operator gives the contact term.

\subsection{Shift and injection kernels}

The paths with $l=n$ in the first output denominator and $l=m$ in the second define the injection sector. Their output energy differences vanish when $\mathbf Q=0$, producing the relaxation-dependent response. The remaining interband paths, together with the contact contribution, define the shift sector. With this separation, the resonant kernels are
\begin{widetext}
\begin{align}
\sigma_{\rm sh}^{abc}(\mathbf q_1,\mathbf q_2)
={}&
-\frac{\ii\pi e^3}{\hbar^2\omega^2}
\sum_{nm\mathbf k}
f_{n_1m}\delta(\omega_{n_1m}-\omega)v^b_{n_1m}
\Bigg[
\sum_{l\ne n}
\frac{v^a_{l_2n_1}v^c_{ml_2}}
{\omega_{n_1l_2}-2\ii\eta}
-
\sum_{l\ne m}
\frac{v^c_{l_{12}n_1}v^a_{ml_{12}}}
{\omega_{l_{12}m}-2\ii\eta}
+
w^{ac}_{mn_1}(-2\mathbf Q,-\mathbf q_2)
\Bigg]
\nonumber\\
&+
\{b,\mathbf q_1\leftrightarrow c,\mathbf q_2\}^{*},
\label{eq:Ash}
\\
\sigma_{\rm inj}^{abc}(\mathbf q_1,\mathbf q_2)
={}&
-\frac{\ii\pi e^3}{\hbar^2\omega^2}
\sum_{nm\mathbf k}
f_{n_1m}\delta(\omega_{n_1m}-\omega)v^b_{n_1m}
\left[
\frac{v^a_{n_2n_1}v^c_{mn_2}}
{\omega_{n_1n_2}-2\ii\eta}
-
\frac{v^c_{m_{12}n_1}v^a_{mm_{12}}}
{\omega_{m_{12}m}-2\ii\eta}
\right]
\nonumber\\
&+
\{b,\mathbf q_1\leftrightarrow c,\mathbf q_2\}^{*}.
\label{eq:Ainj}
\end{align}
\end{widetext}
The final term denotes the complex conjugate of the entire preceding expression after interchanging $(b,\mathbf q_1)$ and $(c,\mathbf q_2)$, including the prefactor and denominators. The common factor $\omega^{-2}$ converts the two vector-potential amplitudes to electric-field amplitudes.

For the interband output denominators, $\eta\to0^+$. The intraband denominators are instead retained at a finite rate, with the phenomenological identification $2\eta=\tau^{-1}$. This relaxation rate is distinct from the numerical regularization used near band degeneracies. The kernels retain the resonant optical poles; nonresonant principal-value optical contributions are not included.

When $\mathbf q_1=\mathbf q_2$, one has $\mathbf Q=0$, and the response reduces to the single-plane-wave photon-drag geometry. Allowing $\mathbf q_1\ne\mathbf q_2$ additionally retains the finite-$\mathbf Q$ current generated by interference between optical components. Expanding Eqs.~\eqref{eq:Ash} and \eqref{eq:Ainj} in $\mathbf q$ and $\mathbf Q$ therefore yields the phase- and intensity-gradient coefficients within the same resonant framework.

\section{Photon drag and the response to spatial intensity variations}
\label{sec:kinematics}

The phase- and intensity-gradient responses follow from two independent derivatives of the same conductivity. Using the definitions
\begin{equation}
\mathbf q=\frac{\mathbf q_1+\mathbf q_2}{2},
\qquad
\mathbf Q=\frac{\mathbf q_1-\mathbf q_2}{2},
\end{equation}
we write
\begin{equation}
\mathbf q_1=\mathbf q+\mathbf Q,
\qquad
\mathbf q_2=\mathbf q-\mathbf Q.
\end{equation}
The conductivity expanded about $\mathbf q=\mathbf Q=0$ is
\begin{align}
\sigma_X^{abc}(\mathbf q+\mathbf Q,\mathbf q-\mathbf Q)
={}&
\sigma_{X,(0)}^{abc}
+q_d\xi_X^{abcd}
+Q_d\zeta_X^{abcd}
\nonumber\\
&+O(q^2,Q^2,qQ),
\label{eq:qQexpS}
\end{align}
where $X=\mathrm{sh}$ or $\mathrm{inj}$. The two coefficients are therefore
\begin{align}
\xi_X^{abcd}
&=
\left.
\frac{\partial}{\partial q_d}
\sigma_X^{abc}(\mathbf q,\mathbf q)
\right|_{\mathbf q=0},
\label{eq:xiDefinitionS}
\\
\zeta_X^{abcd}
&=
\left.
\frac{\partial}{\partial Q_d}
\sigma_X^{abc}(\mathbf Q,-\mathbf Q)
\right|_{\mathbf Q=0}.
\label{eq:zetaDefinitionS}
\end{align}
Equivalently,
\begin{align}
\xi_X^{abcd}
&=
\left.
\left(\partial_{q_{1,d}}+\partial_{q_{2,d}}\right)
\sigma_X^{abc}
\right|_{\mathbf q_1=\mathbf q_2=0},
\\
\zeta_X^{abcd}
&=
\left.
\left(\partial_{q_{1,d}}-\partial_{q_{2,d}}\right)
\sigma_X^{abc}
\right|_{\mathbf q_1=\mathbf q_2=0}.
\end{align}
Thus, $\xi_X$ is obtained by changing the two optical wave vectors together, whereas $\zeta_X$ is obtained by changing them in opposite directions. These choices define the derivatives; they do not require a general optical field to contain only either type of Fourier pair.

For fixed polarization, the Fourier identities derived above give
\begin{align}
J_{a,X}^{(\xi)}(\mathbf r)
&=
\xi_X^{abcd}e_be_c^*
A^2\partial_d\phi,
\\
J_{a,X}^{(\zeta)}(\mathbf r)
&=
-\frac{\ii}{2}
\zeta_X^{abcd}e_be_c^*
\partial_dA^2.
\label{eq:realspaceS}
\end{align}
For a single plane wave,
$\mathcal E(\mathbf r)=A_0e^{\ii\mathbf q_0\cdot\mathbf r}$,
the intensity is uniform and
\begin{equation}
J_{a,X}^{(\xi)}
=
\xi_X^{abcd}e_be_c^*A_0^2q_{0,d},
\qquad
J_{a,X}^{(\zeta)}=0.
\end{equation}
This is the plane-wave photon-drag limit. A spatially varying field can instead contain Fourier pairs with $\mathbf q_1\ne\mathbf q_2$. Their products vary as $e^{2\ii\mathbf Q\cdot\mathbf r}$, and both contributions in Eq.~\eqref{eq:realspaceS} can be present.

The factor $-\ii$ in the intensity-gradient term is consistent with a real physical current. The conductivity satisfies
\begin{equation}
\sigma_X^{abc}(\mathbf q_1,\mathbf q_2)
=
\left[
\sigma_X^{acb}(\mathbf q_2,\mathbf q_1)
\right]^*.
\end{equation}
Since exchanging $\mathbf q_1$ and $\mathbf q_2$ leaves $\mathbf q$ unchanged but reverses $\mathbf Q$, its first-order coefficients obey
\begin{equation}
\xi_X^{abcd}
=
\left[\xi_X^{acbd}\right]^*,
\qquad
\zeta_X^{abcd}
=
-\left[\zeta_X^{acbd}\right]^*.
\label{eq:gradientRealityS}
\end{equation}
Consequently, $\xi_X^{abcd}e_be_c^*$ is real and
$\zeta_X^{abcd}e_be_c^*$ is purely imaginary, so both currents in Eq.~\eqref{eq:realspaceS} are real.

\section{Expansion in the average optical wave vector}
\label{sec:plane}
\label{sec:xi}

To obtain $\xi_X$, set $\mathbf q_1=\mathbf q_2=\mathbf q$ in Eqs.~\eqref{eq:Ash} and \eqref{eq:Ainj}. It is convenient to use the midpoint of the initial and final electronic momenta as the integration variable:
\begin{equation}
n_+=(n,\mathbf k+\mathbf q/2),
\qquad
m_-=(m,\mathbf k-\mathbf q/2).
\end{equation}
The optical transition is then $m_-\to n_+$. In this geometry, the two intraband output energy differences vanish for every $\mathbf q$. The injection bracket therefore reduces to
\begin{equation}
\frac{
(v_{n_+}^a-v_{m_-}^a)v^c_{m_-n_+}
}{
-2\ii\eta
},
\label{eq:injplane}
\end{equation}
where $v_{n_+}^a=v_n^a(\mathbf k+\mathbf q/2)$ and
$v_{m_-}^a=v_m^a(\mathbf k-\mathbf q/2)$.

The interband sums in the shift bracket can be expressed using the usual generalized derivative. With
$\mathcal A_n^a=\ii\langle u_n|\partial_{k_a}u_n\rangle$,
define
\begin{equation}
(v^c_{m_-n_+})_{;k_a}
=
\partial_{k_a}v^c_{m_-n_+}
-\ii(\mathcal A_{m_-}^a-\mathcal A_{n_+}^a)
v^c_{m_-n_+},
\label{eq:covariant}
\end{equation}
where $\mathbf q$ is held fixed. Differentiating the eigenvalue equation gives
\begin{equation}
\langle u_l|\partial_{k_a}u_n\rangle
=
\frac{v_{ln}^a}{\omega_{nl}},
\qquad l\ne n.
\label{eq:HF}
\end{equation}
Applying this identity to the bra and ket of the optical matrix element, and retaining the derivative of the velocity operator, gives
\begin{align}
(v^c_{m_-n_+})_{;k_a}
={}&
w^{ac}_{m_-n_+}(\mathbf 0,-\mathbf q)
+
\sum_{l\ne n}
\frac{v^c_{m_-l_+}v^a_{l_+n_+}}
{\omega_{n_+l_+}}
\nonumber\\
&-
\sum_{l\ne m}
\frac{v^a_{m_-l_-}v^c_{l_-n_+}}
{\omega_{l_-m_-}}.
\label{eq:sumruleq}
\end{align}
Here the contact term is identified with the operator derivative for the local Hamiltonian defined above. Equation~\eqref{eq:sumruleq} uses the zero-broadening interband denominators; replacing them by numerically broadened denominators is a separate approximation.

The plane-wave conductivities consequently become
\begin{align}
\sigma_{\rm sh}^{abc}(\mathbf q,\mathbf q)
={}&
-\frac{\ii\pi e^3}{\hbar^2\omega^2}
\sum_{nm\mathbf k}
f_{n_+m_-}\delta(\omega_{n_+m_-}-\omega)
\nonumber\\
&\times
\left[
v^b_{n_+m_-}(v^c_{m_-n_+})_{;k_a}
-
v^c_{m_-n_+}(v^b_{n_+m_-})_{;k_a}
\right],
\label{eq:planesh}
\\
\sigma_{\rm inj}^{abc}(\mathbf q,\mathbf q)
={}&
\frac{\pi e^3}{\hbar^2\omega^2\eta}
\sum_{nm\mathbf k}
f_{n_+m_-}\delta(\omega_{n_+m_-}-\omega)
\nonumber\\
&\times
(v_{n_+}^a-v_{m_-}^a)
v^b_{n_+m_-}v^c_{m_-n_+}.
\label{eq:planeinj}
\end{align}
The prefactor in Eq.~\eqref{eq:planeinj} includes both exchanged optical contributions. We retain the convention $2\eta=\tau^{-1}$ used in the resonant kernels.

We now expand each factor to first order in $\mathbf q$. For fully occupied and empty bands, the occupation derivative vanishes. The transition energy and velocity difference satisfy
\begin{align}
\left.\partial_{q_d}\omega_{n_+m_-}\right|_0
&=\frac{v_n^d+v_m^d}{2},
\\
\left.
\partial_{q_d}(v_{n_+}^a-v_{m_-}^a)
\right|_0
&=
\frac{\partial_{k_d}v_n^a+\partial_{k_d}v_m^a}{2}.
\end{align}
Writing
$\delta_{nm}=\delta(\omega_{nm}-\omega)$ and
$\delta'_{nm}=\delta'(\omega_{nm}-\omega)$,
the resonance derivative is
\begin{equation}
\left.
\partial_{q_d}\delta(\omega_{n_+m_-}-\omega)
\right|_0
=
\frac{v_n^d+v_m^d}{2}\delta'_{nm}.
\end{equation}
The prime denotes differentiation with respect to the argument of the delta function.

Applying the product rule gives
\begin{align}
\xi_{\rm sh}^{abcd}
={}&
-\frac{\ii\pi e^3}{\hbar^2\omega^2}
\sum_{nm\mathbf k}f_{nm}
\Bigg\{
\frac{v_n^d+v_m^d}{2}\delta'_{nm}
\nonumber\\
&\quad\times
\left[
v_{nm}^b(v_{mn}^c)_{;k_a}
-v_{mn}^c(v_{nm}^b)_{;k_a}
\right]
\nonumber\\
&+
\delta_{nm}
\left.
\partial_{q_d}
\left[
v^b_{n_+m_-}(v^c_{m_-n_+})_{;k_a}
-
v^c_{m_-n_+}(v^b_{n_+m_-})_{;k_a}
\right]
\right|_0
\Bigg\},
\label{eq:xish}
\\
\xi_{\rm inj}^{abcd}
={}&
\frac{\pi e^3}{\hbar^2\omega^2\eta}
\sum_{nm\mathbf k}f_{nm}
\Bigg\{
\frac{v_n^d+v_m^d}{2}\delta'_{nm}
(v_n^a-v_m^a)v_{nm}^bv_{mn}^c
\nonumber\\
&+
\frac{\delta_{nm}}{2}
(\partial_{k_d}v_n^a+\partial_{k_d}v_m^a)
v_{nm}^bv_{mn}^c
\nonumber\\
&+
\delta_{nm}(v_n^a-v_m^a)
\left.
\partial_{q_d}
\left[v^b_{n_+m_-}v^c_{m_-n_+}\right]
\right|_0
\Bigg\}.
\label{eq:xiinj}
\end{align}
The terms proportional to $\delta'_{nm}$ describe the change in the optical resonance. In Eq.~\eqref{eq:xiinj}, the next two terms describe changes in the electron--hole velocity difference and the optical transition matrix elements. In Eq.~\eqref{eq:xish}, the remaining derivative acts on both the optical matrix elements and their generalized derivatives. No derivative of the intraband output denominator appears because that denominator is independent of $\mathbf q$.

The matrix-element derivatives are evaluated by differentiating their initial and final Bloch states and, where applicable, the velocity operator. Derivatives of the complete products in Eqs.~\eqref{eq:xish} and \eqref{eq:xiinj} are independent of the arbitrary phases of the Bloch states.

The midpoint convention used here is related to the initial-momentum convention of Eqs.~\eqref{eq:Ash} and \eqref{eq:Ainj} by
$\mathbf k_{\rm initial}=\mathbf k-\mathbf q/2$.
The corresponding integrand derivatives differ by a total derivative with respect to $\mathbf k$, which integrates to zero over the periodic Brillouin zone. The integrated response is therefore unchanged, although individual contributions may be grouped differently.

\section{Expansion in the optical wave-vector difference}
\label{sec:zeta}

To obtain $\zeta_X$, set $\mathbf q_1=\mathbf Q$ and
$\mathbf q_2=-\mathbf Q$. The electronic states appearing in the resonant kernels are then
\begin{equation}
n_1=(n,\mathbf k+\mathbf Q),\qquad
l_2=(l,\mathbf k-\mathbf Q),\qquad
l_{12}=(l,\mathbf k+2\mathbf Q).
\end{equation}
The positive-frequency optical transition remains $m\to n_1$. For fully occupied and empty bands,
\begin{equation}
\left.
\partial_{Q_d}
\left[
f_{n_1m}\delta(\omega_{n_1m}-\omega)
\right]
\right|_0
=
f_{nm}v_n^d\delta'_{nm}.
\label{eq:Qresonance}
\end{equation}

Unlike the plane-wave case, the intraband output energy differences now depend on $\mathbf Q$:
\begin{align}
\omega_{n_1n_2}
&=
\frac{\varepsilon_n(\mathbf k+\mathbf Q)
-\varepsilon_n(\mathbf k-\mathbf Q)}{\hbar}
\nonumber\\
&=
2Q_dv_n^d+O(Q^3),
\\
\omega_{m_{12}m}
&=
\frac{\varepsilon_m(\mathbf k+2\mathbf Q)
-\varepsilon_m(\mathbf k)}{\hbar}
\nonumber\\
&=
2Q_dv_m^d+O(Q^2).
\label{eq:Qintraband}
\end{align}
The absence of a quadratic term in the first line follows from the symmetric difference between $\mathbf k+\mathbf Q$ and $\mathbf k-\mathbf Q$. The interband denominators similarly satisfy
\begin{align}
\omega_{n_1l_2}
&=
\omega_{nl}+Q_d(v_n^d+v_l^d)+O(Q^2),
\\
\omega_{l_{12}m}
&=
\omega_{lm}+2Q_dv_l^d+O(Q^2).
\label{eq:Qinterband}
\end{align}

Each numerator and denominator must be differentiated. In particular,
\begin{equation}
\partial_{Q_d}
\frac{1}{\omega_{\alpha\beta}-2\ii\eta}
=
-\frac{\partial_{Q_d}\omega_{\alpha\beta}}
{(\omega_{\alpha\beta}-2\ii\eta)^2}.
\label{eq:Qdenometa}
\end{equation}
For the injection denominators, this gives
\begin{align}
\left.
\partial_{Q_d}
\frac{1}{\omega_{n_1n_2}-2\ii\eta}
\right|_0
&=
-\frac{2v_n^d}{(-2\ii\eta)^2},
\\
\left.
\partial_{Q_d}
\frac{1}{\omega_{m_{12}m}-2\ii\eta}
\right|_0
&=
-\frac{2v_m^d}{(-2\ii\eta)^2}.
\label{eq:kinematicDenominatorS}
\end{align}
These derivatives are taken at fixed $\eta>0$, with
$2\eta=\tau^{-1}$. They are absent from $\xi_{\rm inj}$ but contribute to $\zeta_{\rm inj}$.

The numerator derivatives follow from the ordinary product rule. For example,
\begin{align}
&
\left.
\partial_{Q_d}
\left[
v^a_{l_2n_1}v^b_{n_1m}v^c_{ml_2}
\right]
\right|_0
\nonumber\\
&=
\left.(\partial_{Q_d}v^a_{l_2n_1})\right|_0
v_{nm}^bv_{ml}^c
+
v_{ln}^a
\left.(\partial_{Q_d}v^b_{n_1m})\right|_0
v_{ml}^c
\nonumber\\
&\quad+
v_{ln}^av_{nm}^b
\left.(\partial_{Q_d}v^c_{ml_2})\right|_0.
\label{eq:Qproduct1}
\end{align}
Individual matrix-element derivatives depend on the phases chosen for the Bloch states, but these phase-dependent terms cancel in the complete product. The contact product is differentiated in the same way,
\begin{align}
&
\left.
\partial_{Q_d}
\left[
w^{ac}_{mn_1}(-2\mathbf Q,\mathbf Q)
v^b_{n_1m}
\right]
\right|_0
\nonumber\\
&=
\left.
\partial_{Q_d}
w^{ac}_{mn_1}(-2\mathbf Q,\mathbf Q)
\right|_0
v_{nm}^b
+
w^{ac}_{mn}
\left.
\partial_{Q_d}v^b_{n_1m}
\right|_0.
\label{eq:Qcontactproduct}
\end{align}
The first derivative includes the dependence of the contact operator on both momentum arguments as well as the dependence of its electronic states. A vanishing contact matrix element at $\mathbf Q=0$ does not, by itself, imply a vanishing derivative.

Combining the resonance, numerator, denominator, and contact derivatives yields
\begin{widetext}
\begin{align}
\zeta_{\rm sh}^{abcd}
={}&
-\frac{\ii\pi e^3}{\hbar^2\omega^2}
\sum_{nm\mathbf k}f_{nm}
\Bigg\{
v_n^d\delta'_{nm}
v_{nm}^b(v_{mn}^c)_{;k_a}
\nonumber\\
&+
\delta_{nm}\sum_{l\ne n}
\left[
\frac{
\left.
\partial_{Q_d}
\left[
v^a_{l_2n_1}v^b_{n_1m}v^c_{ml_2}
\right]
\right|_0
}{\omega_{nl}}
-
\frac{
(v_n^d+v_l^d)v_{ln}^av_{nm}^bv_{ml}^c
}{\omega_{nl}^2}
\right]
\nonumber\\
&-
\delta_{nm}\sum_{l\ne m}
\left[
\frac{
\left.
\partial_{Q_d}
\left[
v^a_{ml_{12}}v^c_{l_{12}n_1}v^b_{n_1m}
\right]
\right|_0
}{\omega_{lm}}
-
\frac{
2v_l^dv_{ml}^av_{ln}^cv_{nm}^b
}{\omega_{lm}^2}
\right]
\nonumber\\
&+
\delta_{nm}
\left.
\partial_{Q_d}
\left[
w_{mn_1}^{ac}(-2\mathbf Q,\mathbf Q)
v_{n_1m}^b
\right]
\right|_0
\Bigg\}
-\{b\leftrightarrow c\}^{*},
\label{eq:zetash}
\\
\zeta_{\rm inj}^{abcd}
={}&
-\frac{\ii\pi e^3}{\hbar^2\omega^2}
\sum_{nm\mathbf k}f_{nm}
\Bigg\{
\frac{v_n^d\delta'_{nm}}{-2\ii\eta}
(v_n^a-v_m^a)v_{nm}^bv_{mn}^c
\nonumber\\
&+
\frac{\delta_{nm}}{-2\ii\eta}
\left.
\partial_{Q_d}
\left[
v^a_{n_2n_1}v^b_{n_1m}v^c_{mn_2}
-
v^a_{mm_{12}}v^c_{m_{12}n_1}v^b_{n_1m}
\right]
\right|_0
\nonumber\\
&-
\frac{2\delta_{nm}}{(-2\ii\eta)^2}
(v_n^av_n^d-v_m^av_m^d)v_{nm}^bv_{mn}^c
\Bigg\}
-\{b\leftrightarrow c\}^{*}.
\label{eq:zetainj}
\end{align}
\end{widetext}
All unshifted quantities are evaluated at $\mathbf k$. The final term subtracts the exchanged complex conjugate of the entire preceding expression, including its prefactor. This minus sign follows because exchanging the optical components reverses $\mathbf Q$.

In Eq.~\eqref{eq:zetash}, the terms proportional to $\delta'_{nm}$ arise from the resonance shift, the numerator derivatives describe changes in the optical and output-current matrix elements, and the squared denominators arise from the energy-denominator derivatives. The last term differentiates the contact contribution. Equation~\eqref{eq:zetainj} contains the corresponding resonance, numerator, and intraband-denominator terms.

Interband denominators are understood in the zero-broadening limit for nondegenerate bands. The intraband relaxation rate is retained during differentiation and is distinct from any numerical regularization of near-degenerate interband denominators. Together, these expressions show explicitly why the average-wave-vector and wave-vector-difference derivatives contain different contributions, even though both are obtained from the same two-wave-vector conductivity.

\section{First-principles methods and supplemental GeS results}
\label{sec:GeS}

\subsection{Electronic structure and structural interpolation}
The GeS benchmark uses eleven fixed-cell structures with $\lambda=-1.0,-0.8,\ldots,0,\ldots,+0.8,+1.0$.  The $\lambda=+1$ structure is the polar reference, $\lambda=-1$ is its inversion partner, and $\lambda=0$ is the exact centrosymmetric midpoint.  Intermediate structures are generated by linear interpolation and are not separately relaxed, so that $\lambda$ isolates the polar structural coordinate.

The Kohn--Sham calculations are performed with VASP within the PAW/PBE framework using a plane-wave cutoff of $400$~eV and a $10\times10\times1$ $\Gamma$-centered mesh.  The calculations are non-spin-polarized and omit spin--orbit coupling.  Symmetry reduction is disabled along the interpolation so that every structure is sampled on the same full $k$ mesh.  The Wannierization uses 32 Kohn--Sham bands and 16 Wannier functions initialized from Ge and S $s,p$ orbitals; ten bands are occupied in the non-spin-polarized calculation.  The dense response integrals shown below use a $140\times140\times1$ mesh.

\subsection{Wannier-subspace representation of the response vertices}
The response calculation uses the first-principles-derived Wannier Hamiltonian $H^{W}(\mathbf k)$, constructed from the Wannier90 tight-binding matrix elements, as its electronic-structure input.  Energies, eigenvectors, ordinary velocities, band Hessians, and the remote-band sums entering the wave-vector derivatives are evaluated from $H^{W}(\mathbf k)$ and its analytic $\mathbf k$ derivatives.  In the long-wavelength representation used numerically,
\begin{equation}
 v^a(\mathbf k)=\frac{1}{\hbar}\partial_{k_a}H^{W}(\mathbf k),
 \qquad
 w^{ac}(\mathbf k)=\partial_{k_a}v^c(\mathbf k)
 =\frac{1}{\hbar}\partial_{k_a}\partial_{k_c}H^{W}(\mathbf k),
\label{eq:implementedvw}
\end{equation}
and derivatives of $w^{ac}$ are evaluated with the same bra--kernel--ket product rule used for the one-photon vertices.  The intrinsic electric-quadrupole and orbital magnetic-dipole moments, the midpoint/band-dispersion pieces, and the output-current derivatives are then assembled from these Wannier-interpolated quantities and the corresponding remote-band sums.  This is the numerical realization of the long-wavelength derivatives derived in Secs.~\ref{sec:density}.

Three distinct numerical scales are kept separate.  The resonant delta function is represented by a Gaussian of width $0.0680$~eV.  The steady-state injection current uses a phenomenological lifetime $\tau=100$~fs, corresponding to the analytic convention $2\eta=\tau^{-1}$ in Secs.~\ref{sec:density}--\ref{sec:zeta} (equivalently $\hbar\eta=3.29$~meV).  Near-degenerate internal virtual-state denominators in the shift/virtual-state sector are stabilized independently with $\eta_{\rm reg}=0.010$~eV through the real kernel $R_{\eta_{\rm reg}}(\Delta)=\Delta/(\Delta^2+\eta_{\rm reg}^2)$ and its exact derivative.  The parameter $\eta_{\rm reg}$ is a numerical regularizer and is neither the injection relaxation rate nor the Gaussian optical broadening. For the figures we report an effective-three-dimensional normalization using an active monolayer thickness $t_{\rm eff}=2.6$~\AA, consistent with the convention used for comparison with conventional monolayer BPVE calculations.  The plotted rank-four coefficients therefore have units $\mu\mathrm{A}\,\mathrm{\AA}/\mathrm{V}^2$; multiplying by $q_d$ for $\xi$ or $Q_d$ for $\zeta$ in \AA$^{-1}$ gives a second-order conductivity in $\mu\mathrm{A}/\mathrm{V}^2$.  Quantitative comparison to other two-dimensional calculations requires matching the thickness, optical-field convention, and spin multiplicity.

% \subsection{Point-group constraints, tensor components, and inversion parity}
% For each interpolated structure, the crystallographic point-group operations are evaluated in the same Cartesian axes used for the response tensors.  The rank-four coefficients $\xi_X^{abcd}$ and $\zeta_X^{abcd}$ are constrained under the complete point group, and the components displayed in the main text and below are selected from the symmetry-allowed set.  The inversion relation between $+\lambda$ and $-\lambda$ is then used as an independent numerical check, not as the only symmetry criterion.

% Figure~\ref{fig:S_heatmaps} collects four representative large-magnitude symmetry-allowed components from each response family over the calculated photon-energy window.  The rows correspond to $\xi_{\rm sh}$, $\zeta_{\rm sh}$, $\xi_{\rm inj}$, and $\zeta_{\rm inj}$, and every panel title identifies the plotted tensor component and its symmetry-allowed real or imaginary part.  These data supplement the representative components displayed in the main text without using inversion parity as the component-selection rule.  Within numerical accuracy, the displayed finite-wave-vector coefficients are even under $\lambda\rightarrow-\lambda$ and remain finite at the centrosymmetric structure.  In particular, the strong shift-response features near $\lambda=0$ are not tied to a finite polar distortion.

\begin{figure}[p]
\centering
\includegraphics[width=0.98\textwidth]{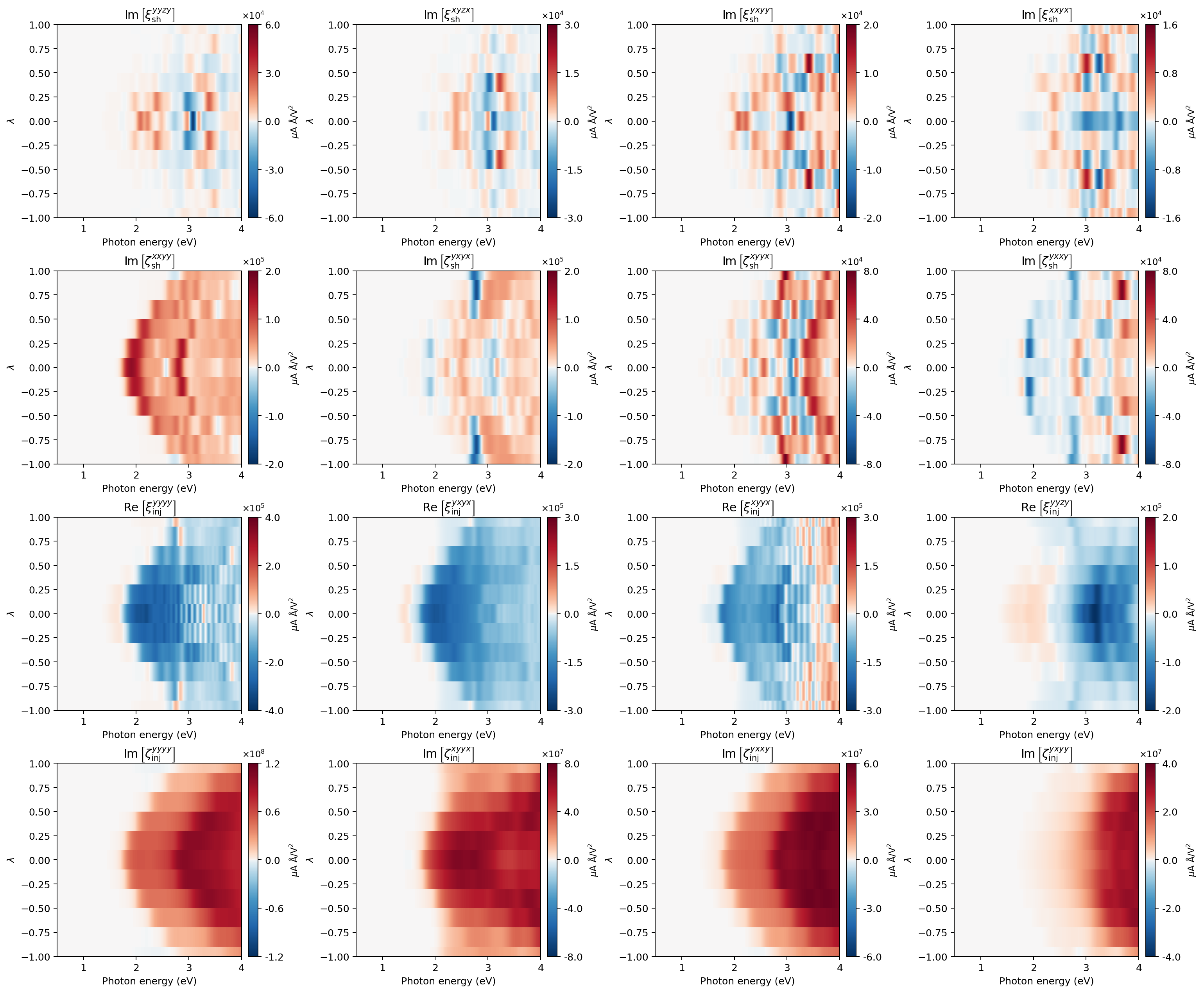}
\caption{Supplemental survey of symmetry-allowed tensor components for monolayer GeS.  The four rows show selected large-magnitude components of $\xi_{\rm sh}$, $\zeta_{\rm sh}$, $\xi_{\rm inj}$, and $\zeta_{\rm inj}$, respectively, as functions of photon energy and structural coordinate $\lambda$.  Each panel title identifies the tensor component and the plotted real or imaginary part.  Color bars report effective-three-dimensional coefficients in $\mu\mathrm{A}\,\mathrm{\AA}/\mathrm{V}^{2}$.}
\label{fig:S_heatmaps}
\end{figure}

\clearpage

% REVTeX 4.2 automatically reinserts the main .bbl at the end of each
% thebibliography environment. Disable that insertion locally for the
% independent Supplemental Material references. Also prevent a second
% LastBibItem label (REVTeX otherwise writes it once per bibliography).
\begingroup
\makeatletter
\let\auto@bib@innerbib\@empty
\let\original@label\label
\def\label#1{%
  \def\temp@label{#1}%
  \def\temp@lastbib{LastBibItem}%
  \ifx\temp@label\temp@lastbib\else\original@label{#1}\fi
}
\makeatother

\endgroup

\end{document}